# Enhancement of a Text-Independent Speaker Verification System by using Feature Combination and Parallel-Structure Classifiers

Kerlos Atia Abdalmalak[1,2], Ascension Gallardo-Antolín[2]

[1]Electrical Engineering Department, Aswan University, Aswan 81542, Egypt
[2]Signal Theory and Communications Department, Carlos III University, Leganes, Madrid 28911, Spain

Kerlos@tsc.uc3m.es, gallardo@tsc.uc3m.es

*Abstract*— Speaker Verification (SV) systems involve mainly two individual stages: feature extraction and classification. In this paper, we explore these two modules with the aim of improving the performance of a speaker verification system under noisy conditions. On the one hand, the choice of the most appropriate acoustic features is a crucial factor for performing robust speaker verification. The acoustic parameters used in the proposed system are: Mel Frequency Cepstral Coefficients (MFCC), their first and second derivatives (Deltas and Delta-Deltas), Bark Frequency Cepstral Coefficients (BFCC), Perceptual Linear Predictive (PLP), and Relative Spectral Transform - Perceptual Linear Predictive (RASTA-PLP). In this paper, a complete comparison of different combinations of the previous features is discussed. On the other hand, the major weakness of a conventional Support Vector Machine (SVM) classifier is the use of generic traditional kernel functions to compute the distances among data points. However, the kernel function of an SVM has great influence on its performance. In this work, we propose the combination of two SVM-based classifiers with different kernel functions: Linear kernel and Gaussian Radial Basis Function (RBF) kernel with a Logistic Regression (LR) classifier. The combination is carried out by means of a parallel structure approach, in which different voting rules to take the final decision are considered.

Results show that significant improvement in the performance of the SV system is achieved by using the combined features with the combined classifiers either with clean speech or in the presence of noise. Finally, to enhance the system more in noisy environments, the inclusion of the multiband noise removal technique as a preprocessing stage is proposed.

## I. INTRODUCTION

Through the past years, Speaker Recognition has become one of the most challenging issues in the field of speech technologies, as it might be crucial for many applications such as transaction authentication, banking operations, database access services, military, voice dialing, in-car systems, Healthcare and remote access to computers [1]. Speaker Verification (SV) is possibly the most important task of speaker recognition. Its purpose is to make the decision of agreement or rejection of a user using exclusively his/her voice (i.e. binary problem). Verification systems can be classified into two different categories: text-dependent [2] and text-independent [3]. In the first one, the user is required to say a predefined utterance like a password. In the second one, the system relies only on the voice characteristic of the speaker regardless the spoken text. In other words, there are no restrictions on the utterances said by the user and, as a consequence, a certain degree of mismatch between the data used in the training and testing phases could appear. For this reason, this kind of systems is more challenging than text-dependent ones. In this work, we focus on text-independent SV systems, for which a detailed overview can be found in [4].

Speaker verification systems consist of two main stages: feature extraction and classification. Acoustic features are parametric representations of speech waveforms and must have high discriminative capabilities. Several works in the literature have addressed the problem of obtaining suitable features for SV, as for example in [5] in which the conventional Mel Frequency Cepstral Coefficients (MFCC) are substituted by the so-called dynamic MFCC parameters. In [6] there is a complete comparison between five different acoustic parameterizations: MFCC, Modified MFCC, Bark Frequency Cepstral Coefficients (BFCC), Revised Perceptual Linear Predictive (PLP), and Linear Prediction Cepstral Coefficients (LPCC) showing that MFCC achieves the highest accuracy compared to the other feature sets. Some other trends in feature extraction can be found in [7] which make use of information that is not contained in cepstral parameters. In [8] instead of using only one feature set, the accuracy of a speaker recognition system is improved by studying different combinations of several complementary features.

The aim of the classifier in the speaker verification system is to compare the given features of the speech utterance to two different models: the claimed speaker model and the impostor model, which can be interpreted as the acoustic pattern representing all other speakers.

In recent years, several types of classifiers have been proposed for speaker recognition, such as Artificial Neural Networks (ANN) [9], [10], [11], Vector Quantization based Probabilistic Neural Network (VQ-PNN) [9], Gaussian Mixture Models (GMM) [12], and Support Vector Machines (SVM) [4], [13], being the two latter techniques the most extended nowadays.

GMM is a common choice for speaker-related tasks as this type of classifier is able to tackle the temporal nature of the speech signal and its mathematical formulation is well-known. Nevertheless, GMMs usually assume that successive acoustic vectors are uncorrelated and follow a Gaussian (or a mixture of Gaussians) distribution, which might not represent properly the feature distributions. On contrast, ANNs and SVMs do not require strong assumptions about the underlying statistical properties of the input features. ANNs present good classification/discrimination properties, however as they are based on the so-called Empirical Risk Minimization (ERM), the probability of converging to a local minimum is great and they are more prone to overfitting [14]. Another drawback of ANNs which limits its usability for speaker verification is that their performance is poor when the size of the available training data is small, which is a common situation in SV systems.

SVM is one of the highest discriminative classifiers that is based on the construction of a hyperplane or set of hyperplanes in a high dimensional space, which simplifies the classification task. Contrary to ANNs, SVMs are based on Structural Risk Minimization (SRM) and hence, they are able to provide a global and unique solution, resulting in better generalization ability. Other advantages of SVMs over ANNs are that they are more suitable for limited training data and able to deal with high-dimensional input vectors, as they use a subset of training points (called support vectors) in the decision function [14]. In addition, SVMs are versatile because of the possibility of using different kernel functions for the decision. If one kernel does not provide the desired performance, the fusing of different kernels could be required. For example, in [15] three kernel functions are combined to achieve better performance for a speaker verification system. For these reasons, in this paper, SVM is one of the classifiers on which the developed SV system is based.

Another interesting discriminative classifier for SV we have experimented in this work is Logistic Regression (LR) [4]. LR is one of the most popular classifiers in the family of Generalized Linear Model (GLM). The advantage of logistic regression is its probabilistic interpretation. In fact, it is based on the computation of the probability of an event occurrence having been given some previously trained data. This way, it can be used directly for SV to guess if an incoming utterance is related to the claimed user (1) or not (0).

In this paper, a study of different features for text-independent speaker verification task is discussed and a comparison between three classifiers: linear kernel SVM, RBF kernel SVM, and Logistic Regression is shown. The main contributions of this paper are the improvement of the feature extraction stage by using a combination of different parameter sets and the enhancement of the classification stage by combining different classifiers. The paper is organized as follows. Section I gives a general overview of speaker verification systems and their main components. Section II presents the state of the art of techniques for building SV systems based on feature and/or classifier combinations. A brief description of the five feature sets used in this work is presented in section III. Section IV deals with the different classifiers proposed for the system. Results with different feature combinations and experiments regarding the use of only one classifier or a combination of all the three classifiers in clean conditions are detailed in section V. Section VI contains experiments and results achieved by the proposed SV systems in different noisy conditions. Section VII shows a comparison between all systems from the point of view of the execution time to check the ability to use them in applications requiring real-time. Finally, all work is concluded in section VIII with plans for future work.

## II. STATE OF THE ART

In this paper, we consider two main techniques to improve the performance of speaker verification systems:

1. Enhancement of the extracted features, by combining different feature sets with high discriminative capability.
2. Enhancement of the classification stage, by combining classifiers using a parallel structure with different voting rules.

Regarding the feature extraction process, nowadays there is a trend of combining different speech features to get higher performance. This combination helps in covering different aspects such as the difference in the rhythm, pronunciation pattern, accents, intonation style and so on. In this context, the authors in [16] combine Linear Prediction Coding (LPC) with MFCC to increase the performance of a speaker recognition system. In another related work, a combination of Discrete Wavelet Transform (DWT) and Relative Spectral Transform - Perceptual Linear Predictive (RASTA-PLP) is proposed [17]. In [18] the authors also combine DWT with traditional MFCC to build a robust speaker recognition system.

With respect to the classification module, SVM has recently become one of the most common and robust classifiers for speaker verification due to, among the other advantages previously mentioned, its good generalization capability to classify unseen data [19] in addition to its ability for training nonlinear decision boundaries efficiently [20]. The kernel function is the key part of SVM and, in fact, the choice of it could change the learning ability of the classifier. Different kernels, which mean different transformations and characteristics, will provide different accuracy for the classification. Based on the analysis of various kernel functions, it can be expected that the best performance of the classifier comes from the combination of several of these functions resulting in higher generalization and learning capabilities.

In addition to this, the way of combination and the decision rule used to make the final decision have a major effect on the whole system. In this context, in [21] the authors compared the combination of four different kernels to build an SV system: linear, quadratic, polynomial of order 3 and RBF with $\sigma = 1$. The main conclusion extracted from this study was that in any possible case, results of using only one kernel function were worse than those achieved by using the aggregation of all other functions. Therefore, we can hypothesize that the combination of kernel functions will improve the performance of the whole SV system as it avoids the weakness of each one of the individual classifiers and increases their strengths.

There are three common structures for combining multiple classifiers [22]:

1. Serial structure. In this case, each individual classifier is invoked sequentially, with the results of the classifier *N-1* being used as input for the classifier *N* which is following it in the sequence.
2. Parallel structure. The outputs of the individual classifiers are passed to the combiner which makes the final decision by using some rules.
3. Hierarchical structure. In this case, the outputs of the individual classifiers are feeding as inputs to the classifier which is the parent node of them.

In addition, there are few 'fixed' rules for making a final decision in the combination of *N* classifiers [23]:

1. AND rule. Speaker utterance *x* is declared as class 1 (verified user) if all classifiers predict it as class 1; otherwise, it will declare as class 0.
2. OR rule. Speaker utterance *x* is declared as class 1 (verified user) if at least one of the classifiers predicts it as class 1; otherwise, it will declare as class 0.
3. Majority vote rule. Speaker utterance *x* is declared as class 1 if the majority of the classifiers predict it as class 1; otherwise, class 0 is declared [24].
4. k-out-of-N rule: Speaker utterance *x* is declared as class 1 if at least *k* of the *N* classifiers predict it as class 1. The previous three rules are special cases of this one with different values of *k*.

In our experiments, we compare the performance of the SV system using the first three rules of combining classifiers.

### III. PROPOSED FEATURE COMBINATION

It is very difficult for the classifier to take raw speech data directly as the input. In fact, this may affect the learning ability of the classifier and dramatically decreases the accuracy of the system. Because of this, the input speech waveforms must be transformed to a set of acoustic features with the following desired characteristics: non-redundancy, dimensionality reduction, and discriminative capabilities.

This transformation stage is called feature extraction. Features must be accurately chosen to lead to better performance of the system. In this paper, we propose the combination of five different feature sets, which are the following.

#### A. Mel-Frequency Cepstral Coefficients

MFCC is one of the most popular parameterizations in speech and speaker recognition tasks. It was developed by Davis and Mermelstein in 1980 [25]. The MFCC extraction method considers (at least, partially) knowledge about how humans hear acoustic signals. MFCC is computed by applying the Discrete Cosine Transform (DCT) on the logarithm of the short-term energy spectrum after converting it using a nonlinear Mel-frequency scale. The idea behinds this frequency warping is to make an approximation to the non-equal sensitivity of the human hearing at different frequencies. In fact, as it can be observed in Fig. 1, the Mel scale does not depend linearly on the normal frequency as it is roughly linear below 1 kHz and logarithmic above this value.

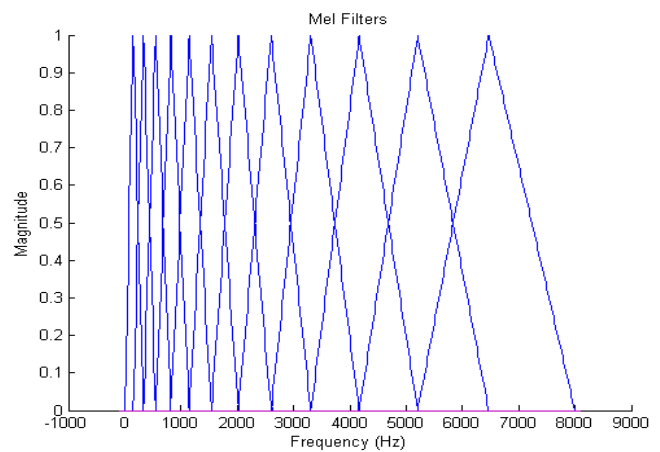

Fig. 1. The Mel frequency scale.

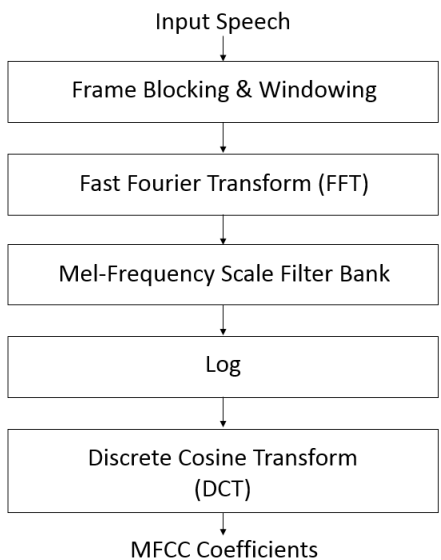

Fig. 2. Extraction of MFCC features from speech signals.

Fig. 2 presents the block diagram of the MFCC extraction process. To transform from the linear frequency to the Mel frequency, eq. (1) is used.

$$Mel = 2595 \log(1 + f/700) \qquad (1)$$

where *f* is the linear frequency (Hz) and *Mel* is the Mel one. In the final step, to obtain the coefficients, the log Mel spectrum is converted back to time by using the DCT.

MFCC is a good and compact representation of the speech signals as it takes into consideration some characteristics of the human auditory system. In fact, the MFCC parameterization is the most commonly used in speech-related tasks. However, its major weak point is its high sensitivity to noise interference (in general, to any kind of mismatch between training and testing conditions) that dramatically decreases the performance of the system. Also, in the computation of MFCC, only the magnitude spectrum of speech signals is used and therefore any potentially relevant information contained in the phase spectrum is ignored [26].

In our experiments, MFCC parameters with different dimensions were obtained from speech utterances. In all cases, feature vectors were extracted on 25 ms Hanning analysis windows, each at 10 ms (the step between successive windows which allows some overlap to the frames). These values are common in feature extraction from speech because a smaller window length will release a low number of samples in the frames which will not be enough to get the reliable information and a larger length of the window will give frequent changes in the information inside the frame. The Mel-cepstral vectors were computed using a triangular Mel-scaled filter bank of 40 filters on the magnitude spectrum of the speech signal.

### B. Deltas and Delta-Deltas of MFCC

Deltas and Delta-Deltas of MFCC are also called differential and acceleration coefficients or dynamic coefficients. The MFCC feature vector (static coefficients) describes only the power spectral envelope of a single frame, but the delta and delta-deltas describe the dynamics information of speech i.e. the behavior of the MFCC trajectories over time.

They are the same dimension as MFCC coefficients; for example, if we have *N* features, we would also get *N* deltas coefficients and *N* delta-deltas coefficients which will combine to give a total feature vector of length 3×*N*. The delta coefficients are calculated using the following equation:

$$D_t = \frac{\sum_{i=1}^{W} i(C_{t+i} - C_{t-i})}{2\sum_{i=1}^{N} i^2} \qquad (2)$$

where $C_t$ represents the MFCC coefficients at the *t-th* frame, $D_t$ is the delta coefficients at *t-th* frame and *W* is the length of the window for computing the delta features.

The Delta-Deltas parameters are calculated in a similar way but with replacing the MFCC coefficients in the previous equation by the delta coefficients.

Dynamic features contribute to improving the robustness of the speaker verification system to noise in comparison with using only MFCC. But on contrary, they also carry additional information, as for example the speech rate, which might not be appropriate for speaker recognition systems (at least when no noise is present).

In our experiments, we have used windows of 9 points (i.e. *W* = 4) to calculate the dynamic parameters from the MFCC ones.

### C. Bark Frequency Cepstral Coefficients

The Bark scale provides a better and motivated scale in comparison to the previous Mel one. Each point on the basilar membrane can be considered as a band pass filter having a variable bandwidth equal to one critical bandwidth [27]. These critical band units are called "Bark". The following equation represents the function which transforms the linear frequency to the Bark frequency:

$$Bark = 6 \sinh^{-1}(f/600) \qquad (3)$$

where *f* and *Bark* are, respectively, the linear (Hz) and Bark frequencies.

After applying the Bark-scaled filter bank on the magnitude spectrum of the speech signal, the remaining conversions to get the final features are the same as MFCC.

As the BFCC extraction process is similar to the MFCC computation excepting for the frequency warping used, BFCC and MFCC basically share the same limitations.

As in the case of MFCC, BFCC parameters with different dimensions were obtained from speech utterances on 25 ms Hanning analysis windows, each at 10 ms. Then a Bark frequency warp was used to get the Bark-cepstral vector by applying a Bark-scale triangular filter bank of 40 filters on the magnitude spectrum of the speech signal.

### D. Perceptual Linear Predictive Coefficients

This method was proposed by Hermansky [28]. It uses three main concepts from the psychophysics of hearing (which improves the performance of the system) to derive an estimate of the auditory spectrum and obtain the PLP coefficients: (1) the critical-band spectral resolution, (2) the equal-loudness curve, and (3) the intensity-loudness exponential law which is known as the cubic-root compression, which is implemented by using eq. (4),

$$S(m) = \left(\sum_{k=0}^{N-1} |X_m(k)|^2 H_m(k)\right)^{0.33} \quad (4)$$

where $X$ is the magnitude spectrum of the speech signal, $H$ is the perceptual filter considered (in this case, the Bark frequency warping is used), $m$ is the index of the filter band which ranges from 0 to 39 (40 filters), $N$ is the number of frames and $S$ is the modified auditory spectrum. This equation is an approximation of the power law of the hearing human system which shows the nonlinear relationship between the intensity of the sound and its perceived loudness.

The complete block diagram of extraction of the PLP features from the speech signal is shown in Fig. 3.

PLP aims to combine MFCC and LPC advantages. Indeed, it is identical to LPC except that the speech spectral characteristics are transformed to match those of the human hearing system [29]. However, PLP analysis presents several drawbacks. On the one hand, the critical band filtering stage may introduce a spectral smoothing with limits its discrimination capabilities for speaker recognition tasks [30]. On the other hand, PLP features are dependent on the whole spectral balance of the formant amplitudes, which is highly sensitive to noise and channel distortions [26].

In our experimentation, 9-dimensional PLP feature vectors were extracted from speech utterances. The analysis window was Hanning of 25 ms length with a step time of 10 ms.

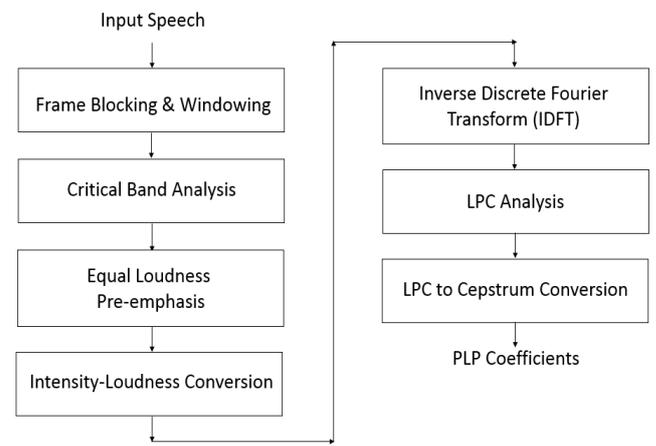

Fig. 3. Extraction of PLP features from speech signals.

### E. Relative Spectral Transform - Perceptual Linear Predictive Coefficients

These features, proposed by Hermansky and Morgan [31], are based on the application of the RASTA band-pass filtering to the PLP features to compensate the distortions caused by linear channels and get robust speaker recognition.

One of the advantages of this filter is the ability to be used either in the log spectral or cepstral domains. It removes any constant offset resulting from static spectral coloration in the speech channel and produces a smoothing over short-term noise variations. Although RASTA-PLP features perform relatively well when there is a mismatch between the train and test conditions, in some occasions, its performance degrades in clean speech scenarios.

RASTA-PLP features were extracted using the same kind of analysis window and step time as in the case of PLP. The RASTA filter was a band-pass filter with a single pole at 0.94. Finally, 9-dimensional RASTA-PLP feature vectors were computed from speech utterances.

## IV. PROPOSED CLASSIFIERS

### A. SVM classifier with different kernels

SVM is one of the most used binary classifiers, which nowadays has become popular for speaker verification tasks [13]. In contrast to traditional methods for speaker verification that separately model the probability distributions of claimed speaker and impostors, SVM discriminates between the different classes by using a set of hyperplanes that satisfy the maximum separation criterion. Fig. 4 shows a 2D example for SVM classifier in which the data are linearly separable.

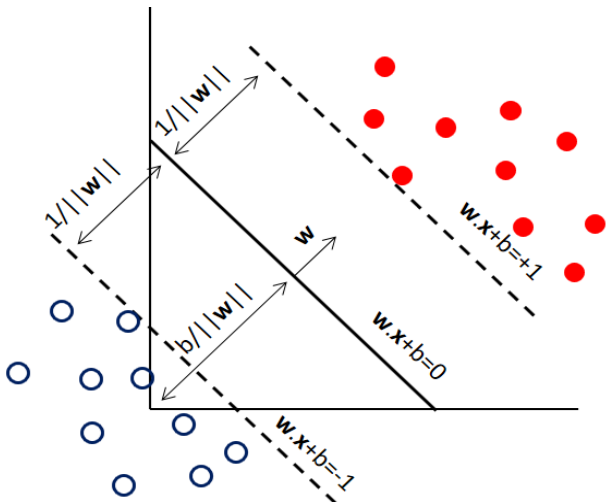

Fig. 4. An example of SVM classifier.

The hyperplanes can be defined by the following equation:

$$\mathbf{w}.\mathbf{x} + b = 0 \qquad (5)$$

where $\mathbf{w}$ is the normal vector to this separating hyperplane, $\mathbf{x}$ is the vector containing the features points and $b$ is a real parameter which determines the offset of the hyperplane from the origin. The perpendicular distance from the hyperplane to the origin is $b/\|\mathbf{w}\|$, where $\|\mathbf{w}\|$ is the Euclidean norm of $\mathbf{w}$.

The goal for the SVM is to find the optimal position of the hyperplane that maximizes the margin between the two dash lines in Fig. 4. This corresponds to minimize $2/\|\mathbf{w}\|$. With the restriction of all samples being correctly classified, the margin is called a hard-margin which achieves the following inequality [32],

$$y_i(\mathbf{w}.\mathbf{x}_i + b) \geq 0 \qquad (6)$$

where $y_i$ is either $1$ or $-1$, indicating the class that the feature vector $x_i$ belongs to.

In the case of non-separable data, another type of margin, called soft-margin, is required. In this case, there is a tradeoff between increasing the size of the margin (i.e. to accept some of the misclassified data in the training phase) and ensuring that all data are classified correctly (which can affect the SVM generalization capability and produce the over-fitting problem). The slack parameter which controls this tradeoff is called $C$. Low values of $C$ imply a low number of training errors but with low generalization capability and vice versa.

When the classes cannot be linearly separated, it is possible to use kernel functions to provide a simple transformation from linear to non-linear spaces. In other words, the original data are mapped and transformed using the kernel function into linearly separable data. In this paper, we propose the use of two combined kernel functions to achieve the optimal performance and enhance the SVM learning ability [33]. The selected two kernels are:

*1) Linear Kernel*

It is the most common and simplest type in SVM. In linear kernel, the function is just the dot product between two feature vectors,

$$K(x_i, x_j) = f(x_i).f(x_j) \qquad (7)$$

where $K(.)$ is the kernel function, $f(.)$ is the mapping function from the input space to another space where the data can be linearly separated, and $x_i$ and $x_j$ are two feature vectors with the same dimension.

*2) Gaussian Radial Basis Function Kernel*

The RBF kernel is equivalent to a linear kernel in an infinite-dimensional feature space, but still easy to compute. It uses the Gaussian function to make the transformation from the nonlinear separation between the two classes to a linear one. It is a special case for the generalized radial basis function and can be expressed by the following equation [14]:

$$K(x_i, x_j) = \exp\left(\frac{-\|x_i - x_j\|^2}{2\sigma^2}\right) \qquad (8)$$

where $x_i$ and $x_j$ are two features vectors with the same dimension, $\|x_i - x_j\|^2$ is the squared Euclidean distance between these two feature vectors and $\sigma$ is the standard deviation which controls the width of the Gaussian radial basis function.

As mentioned before, one of the reasons for selecting SVM as a classifier in this work is because it is well suitable to deal with high dimensional feature vectors as the construction of the hyperplanes in the high dimensional space do not require much computational complexity. A detailed discussion of the computational complexity of SVM can be found in [34]. These high dimensional features may emerge after the combination of several parameter sets as it will be shown in section V.

B. Logistic regression

LR is a special case of GLM which is a large class of statistical models for relating responses to linear combinations of predictor variables. It is widely used in problems with high dimensional settings [35] (what is our case here). LR can be represented by the following equation [36],

$$y_i = G(\mu_i) = \mathbf{x}_i.\boldsymbol{\beta} \qquad (9)$$

where $y_i$ is the predicted value indicating the class that the feature vector $\mathbf{x}_i$ belongs to, $\mathbf{x}_i$ is the feature vector, $\boldsymbol{\beta}$ is a vector of unknown parameters, $G(.)$ is the link function and $\mu_i$ is the expected value of the variable $y_i$, such as $\mu_i = E(y_i)$. Note that "." is the dot product between the two vectors.

The link function can be any differentiable one but it is preferable the use of functions which its inverse link is easily computed such as Poisson, Gamma, inverse Gaussian, and so on. In comparison to the ordinary linear regression which predicts the expected value of a given unknown quantity (the feature vector related to the claimed user or not) as a linear combination of a set of the features, GLM generalizes it by allowing the linear model to be related to the response variable via the link function. So the main difference between them is that the linear regression is a GLM in which its link function is the identity.

The particular LR algorithm [37] considered here is a binomial logistic regression which uses the logit link function as expressed in eq. (10),

$$G(\mu_i) = \log\left(\frac{\mu_i}{1-\mu_i}\right) = \boldsymbol{x}_i \cdot \boldsymbol{\beta} \qquad (10)$$

In the training phase, LR uses this logistic function to find a suitable model by estimating the coefficients of $\boldsymbol{\beta}$ which better fit the current features. In the test phase, LR uses this model for estimating the probability of the incoming feature vectors of belonging to the claimed user or not.

## V. Experiments on Clean Speech

### A. Database

In this paper, the experiments were performed using the English Language Speech Database for Speaker Recognition (ELSDSR) provided by the Department of Informatics and Mathematical Modeling (IMM) at Technical University of Denmark (DTU) [38]. The ELSDSR dataset was designed specifically for speaker recognition purposes by Feng and Hansen. It was recorded in a noise free environment with a fixed microphone containing 22 speakers (12 males, 10 females) from different countries with a high range of ages from 24 to 63. Each speaker uttered 9 paragraphs whose text was taken from NOVA home [39]. This text was provided to capture all the possible pronunciations of English language including vowels, consonants, and diphthongs, etc. [40].

### B. Experimental protocol

The whole dataset was split into two groups: training and testing. The training set contains 154 utterances (7 paragraphs × 22 users), whereas the remaining 44 utterances (2 paragraphs × 22 users) samples are used for testing. The average duration of reading time for the training data is 83 sec and 17.6 sec for testing data.

In order to avoid the problem of overfitting and to select the best parameters of the classifier, it is required to perform cross-validation. It consists of splitting the training set into training and validation subsets and doing performance measurements on the validation subset. In particular, the SVM cost parameter $C$ needs to be optimized by cross-validation, so a 7-fold method is used with a set of values of $C$ = [1/32, 1/16, 1/8, 1/4, 1/2, 1, 2, 4, 8], in such a way that the optimal values which achieve the maximum average performance on the validation set are taken. After selecting the best parameters, whole training data are used for building the final classifier.

The only difference when the RBF kernel is considered with respect to the linear case is that RBF has one extra parameter to be optimized ($\sigma$) and therefore, it is necessary to select the pair of values which achieves the best average performance through cross-validation. The set of $\sigma$ values considered is $\sigma$ = [1/8, 1/4, 1/2, 1, 2, 4, 8, 16, 32].

### C. Performance measurement

In order to compare the performance of the different combinations tested, we need a reliable measurement of success of the system. Many papers in the literature use Detection Error Trade-Off (DET) curves [4] or Receiver Operating Characteristic (ROC) curves [41], [11] for this purpose. Both kind of measures are very closely related, as DET plots the false negative rate (miss detections) against the False Positive Rate (FPR) instead of representing the True Positive Rate (TPR) against FPR (as in the case of ROC), on a non-linear scale (logarithmic or normal deviate scale) instead of a linear one (as in the case of ROC).

Moreover, in order to facilitate the comparison between systems and for the sake of brevity, some scalar measures have been also proposed. Among them, it is worth mentioning the Equal Error Rate (EER) (operating point at which the false acceptance and false rejection probabilities are equal) and the Area Under Curve (AUC) [42]. Note that while EER only refers to the performance of the system in a single point (where miss detections equal false alarms), AUC summarizes the ROC curve for all operating conditions. In this context, recently the work in [43] has shown that the optimization of a speaker verification system with respect to the AUC measure instead of EER is a more robust strategy, especially in noisy conditions. For these reasons, in this paper, we have leaned towards the use of AUC as a measure of the overall performance of the speaker verification system, as in other recent works (see, for example, [44], and [45]).

For computing the AUC of the system first, its ROC is obtained by plotting the TPR (or Recall) as in eq. (11) against the FPR (or 1- specificity) as in eq. (12),

$$TPR = \frac{TP}{TP + FN} \qquad (11)$$

$$FPR = \frac{FP}{FP + TN} \qquad (12)$$

where,
*TP* (True Positives): system accepts a valid user.
*FP* (False Positives): system accepts an impostor as a valid user.
*TN* (True Negatives): system rejects an impostor.
*FN* (False Negatives): system rejects a valid user.

ROC measurements work much better than the traditional accuracy measure (from the point of view of the reliability) which is calculated by dividing the number of the correctly verified users by the total number of testing users [46]. This is due to the unbalanced testing data (2 from positive class and 42 from the negative one). For example, if we have a bad system that classifies any sample as the dominant class (negative class) the accuracy of that system will be around 0.95 but the AUC value will be 0.5. This shows that the AUC metric gives more reliable results in this case.

### D. Results with feature combination using one individual classifier.

This set of experiments on clean speech was carried out for studying the performance of the five types of acoustic parameters initially selected and their combinations. The individual feature sets were: MFCC, first and second derivatives of MFCC (denoted as D+DD, for simplicity), BFCC, PLP and RASTA-PLP (denoted as R-PLP, for simplicity). A publicly available MATLAB implementation was used to generate these features [47].

For brevity, only results with the most successful combinations of the previous feature sets are shown. In summary, Table 1 contains the selected features experimented.

Table 1. Index of the selected combination of features.

| Index of feature combination | Feature sets |
|---|---|
| 1 | MFCC |
| 2 | D+DD |
| 3 | MFCC+D+DD |
| 4 | BFCC |
| 5 | R-PLP |
| 6 | MFCC+BFCC |
| 7 | MFCC+PLP |
| 8 | MFCC+BFCC+PLP |
| 9 | MFCC+BFCC+R-PLP |
| 10 | MFCC+BFCC+PLP+R-PLP |
| 11 | MFCC+D+DD+BFCC |
| 12 | MFCC+D+DD+PLP |
| 13 | MFCC+D+DD+BFCC+PLP+R-PLP |

In the existing literature, results with some of these selected features and their combinations have been reported. In particular:

1. MFCC, BFCC and their combination have been studied in [48], concluding that the combination of both features sets does not increase the performance of the system quality because they characterize the same properties of the speech signal.
2. MFCC, MFCC+D+DD, and R-PLP have been experimented in [49] showing that R-PLP performs better than MFCC even with the MFCC dynamic parameters are also included.

In our case, results are obtained with the acoustic parameters contained in Table 1 when using three different classifiers: linear SVM, RBF SVM, and logistic regression. After a preliminary experimentation in which the optimal feature vector dimensions were chosen, we fixed the length of the acoustic vectors to 10 for MFCC, D+DD, and BFCC and to 9 for PLP and R-PLP.

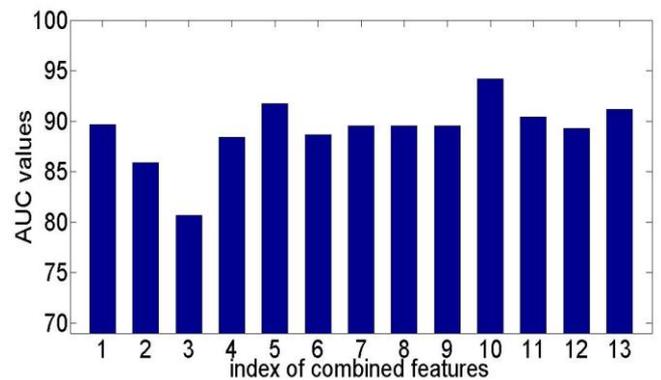
Fig. 5. AUC values for different feature combinations and linear SVM.

Fig. 5 shows the results of the different feature sets considered with the SVM classifier which uses the dot product kernel (linear kernel). As it can be observed, best results come from the combination of MFCC with BFCC, PLP and R-PLP (index 10) which gives an AUC value around 94%. In comparison, the results when all the features are combined (index 13) are slightly lower (around 91%). R-PLP features (index 5) have the second rank of high performance (around 92%). The minimum AUC achieved is 81% which is obtained with MFCC+D+DD (index 3).

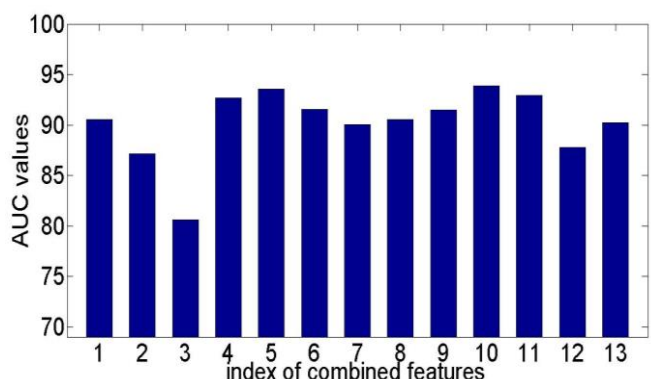
Fig. 6. AUC values for different feature combinations and RBF SVM.

The AUC values obtained with the system based on RBF SVM are shown in Fig. 6. It can be observed, that, in general, these results are similar to the case of the linear kernel.

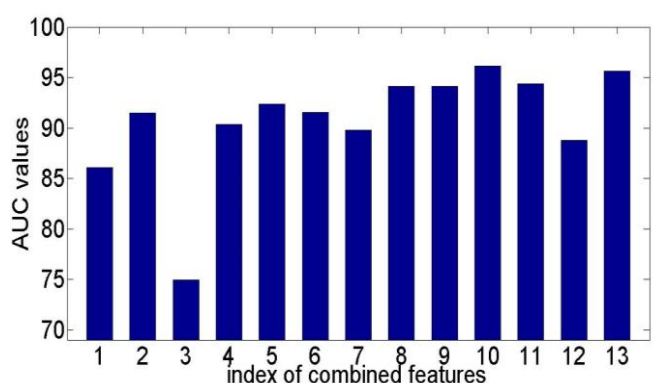
Fig. 7. AUC values for different feature combinations and LR.

Fig. 7 shows the AUC values achieved by the logistic regression-based classifier and different feature combinations. In this case, results are better than those achieved by the SVM classifiers for MFCC+BFCC+PLP+R-PLP (index 10) and MFCC+D+DD+BFCC+PLP+R-PLP (index 13) but worse in some other combinations such MFCC (index 1) and MFCC+D+DD (index 3).

From these experiments, it is possible to draw the following general conclusions:

1. The addition of first and second derivative parameters to the MFCC does not enhance the text-independent speaker verification system, but it dramatically decreases its performance. As mentioned before D and DD try to capture the dynamics of the temporal trajectories of MFCC, which play an important role in phoneme perception. In fact, D and DD have been proved to improve the performance of automatic speech recognition systems, especially in the case of speaker-independence [50]. Traditionally, parameterization modules for speaker recognition systems have been inherited from the speech recognition field, so the use of MFCC and their first and second derivatives is a common practice in SV. However, dynamic features also carry additional information, as for example the speech rate, which can lead to errors in speaker recognition systems (at least when no noise is present). In our opinion, this is the reason for the poor performance of the combination MFCC+D+DD in our system in clean conditions. Similar results have been reported in other works on speaker recognition or related tasks [51], [52].

2. The highest AUC values are achieved with the MFCC+BFCC+PLP+R-PLP parameters. This feature combination seems to be a better representation of the acoustic characteristics of the speakers and gives the highest performance regardless the used classifier.

3. MFCC+BFCC (index 6) works similar to MFCC only for linear and RBF SVM-based classifiers (as the conclusion of [48]). Nevertheless, with logistic regression this combination gives much better results in comparison to MFCC.

4. Regardless the used classifier, R-PLP performs better compared to both MFCC only and MFCC+D+DD. This conclusion agrees with the results reported in [49].

5. No single classifier is clearly considered the best for any combination of the features.

### E. Results with the parallel combination of classifiers.

The previous experiments showed that out of the three classifiers considered, no single classifier is clearly best. However, in general, the group of misclassified samples may not be the same for all of them. Thus, different classifiers may give complementary information and then their combination could be valuable.

In this Section, we present a set of experiments on clean speech carried out for studying the performance of the SV system when the combination of the previous three classifiers is considered. The structure we have used for combining the classifiers is the parallel one. The comparison between the well-known three rules: majority voting (all weights equal), AND and OR rules is shown in the following three figures.

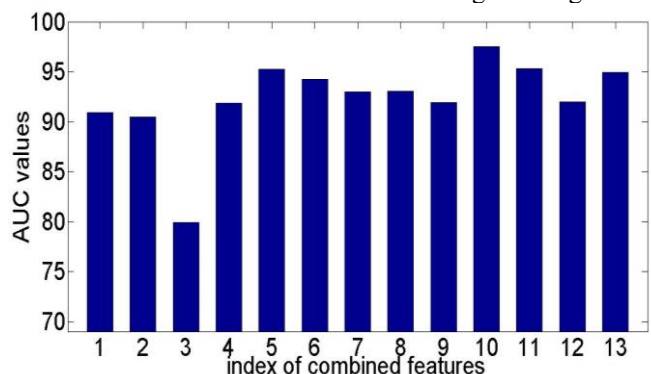
Fig. 8. AUC values for the combination of the three classifiers considered with the majority voting rule.

Fig. 8 shows that the results obtained by the combination of classifiers and the majority voting rule are in most of the cases slightly better than those achieved by the individual classifiers. In the worst case, if one classifier performs badly as MFCC (index 1) or MFCC+D+DD (index 3) in Fig. 7 or MFCC+BFCC (index 6) in Fig. 5, it can be observed that the performance of the combination is medium.

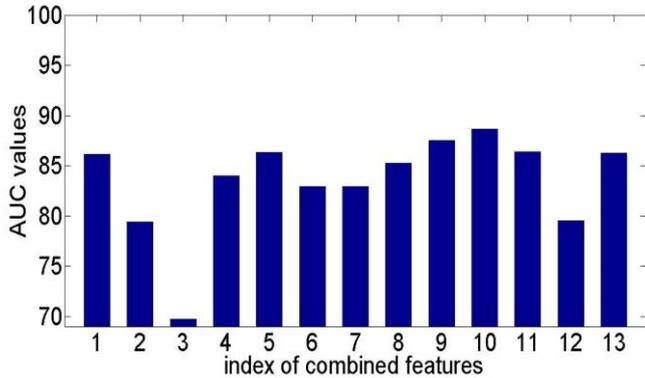
Fig. 9. AUC values for the combination of the three classifiers considered with the AND rule.

Fig. 9 shows the results achieved by using the AND rule. As it can be observed, the performance of the whole system using this rule is not good. This is because the positive class is the minor class in our unbalanced data, and therefore detecting it is harder than the majority class. So this rule increases the probability of misclassifying the samples from this minor class.

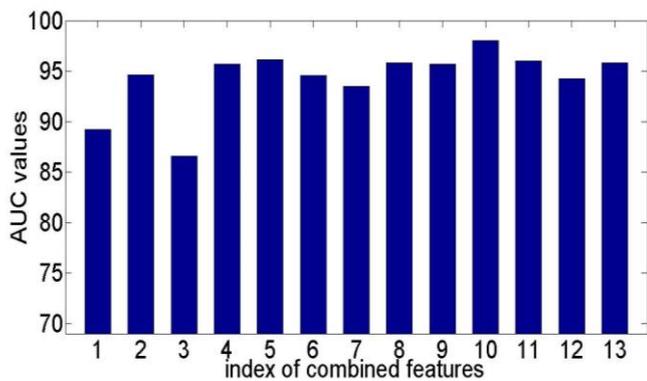
Fig. 10. AUC values for the combination of the three classifiers considered with the OR voting rule.

Fig. 10 contains the AUC values obtained with the OR rule. From these results, we can see that in any possible case, the AUC values achieved with only one classifier are worse than those obtained by using the combination of the three classifiers and the OR rule. This fact suggests that this rule is more efficient because it avoids the weakness of each of the single classifiers and increases their strengths.

From Fig. 10 it can be observed that MFCC+BFCC+PLP+R-PLP (index 10) provides a very high AUC value (around 98%). This result is relatively high in comparison to 94%, 94%, and 96% achieved by linear SVM, RBF SVM, and logistic regression, respectively.

As a general conclusion from the previous six figures, we can see that the feature set composed of MFCC, BFCC with PLP and R-PLP gives the highest performance using any individual classifier or using the combination of classifiers with any rule.

With respect to the classification module, in general, the combination of linear SVM, RBF SVM and LR with the OR rule produces the best results in comparison to the individual classifiers and other combination schemes. The system based on the majority voting rule also achieves satisfactory results in clean speech. Nevertheless, the OR rule is selected for the rest of our experimentation, as it provides in clean conditions an approximately flat performance with high accuracy regardless of the used features. In addition, in a preliminary experimentation with noisy speech, it was observed that the OR rule provides a significantly higher performance than the majority voting, especially for very low Signal-to-Noise Ratios (SNR) values, as for this specific application, the FP error of the individual classifiers is negligible in comparison to the FN one. This fact is also clear from the significant reduction of performance of the system based on the AND rule for any feature combination.

## VI. EXPERIMENTS ON NOISY SPEECH

In this section, the results achieved by the different proposed SV systems under several noise conditions are presented.

### A. First proposed system

Fig. 11 shows the diagram of the overall proposed system which consists of the combination of four feature sets with different dimensions in the feature extraction module and the combination of three classifiers with the OR voting rule to take the final decision of acceptance or rejection in the classification module.

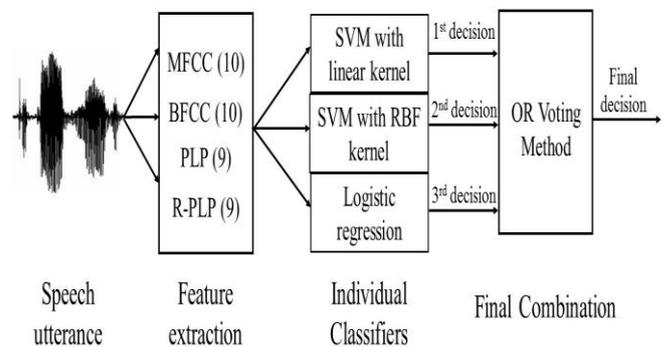
Fig. 11. The first proposed speaker verification system.

For comparative purposes, the performance "at the same dimension of features" of other three basic systems is also presented. These reference systems are: MFCC only (denoted as "Basic 1"), MFCC+D+DD (denoted as "Basic 2") and MFCC+D+DD+BFCC+PLP+R-PLP (denoted as "Basic 3"). In order to extend the comparison, in addition to the previous systems which include MFCC in the feature vectors, another three systems involving PLP parameters are also considered. These PLP-based reference systems are: PLP+BFCC (denoted as "Basic 4"), PLP+R-PLP (denoted as "Basic 5"), and PLP+BFCC+R-PLP (denoted as "Basic 6"). All the reference systems use the linear kernel SVM classifier. Note that in spite of the poor performance of MFCC+D+DD for clean speech, in this set of experiments in noisy conditions, we decided to augment MFCC with dynamic information by including the corresponding first and second derivatives in the second and third reference systems, as several works have shown that this parameterization approach improves the robustness of the system to noise in comparison with using only MFCC [53].

Table 2 shows a detailed comparison between the results in noisy scenarios using the proposed system shown in Fig. 11 and the six basic systems. For doing these experiments, the database was artificially contaminated with eight different noises (airport, babble, car, exhibition, restaurant, street, subway, and train) at three different SNRs: 20, 15 and 10 dB. The different classifiers were trained with a clean speech in all cases, whereas the test was performed in noisy conditions.

Table 2. AUC values for various noise types and SNRs for the first proposed system.

| Noise type | System | SNR values | | |
|---|---|---|---|---|
| | | 20 dB | 15 dB | 10 dB |
| Noise 1 (airport) | Basic 1 | 61.5 | 59.1 | 57.0 |
| | Basic 2 | 59.7 | 55.3 | 54.8 |
| | Basic 3 | 60.7 | 58.8 | 56.6 |
| | Basic 4 | 61.7 | 58.8 | 56.0 |
| | Basic 5 | 57.0 | 55.1 | 54.5 |
| | Basic 6 | 58.0 | 57.7 | 57.2 |
| | Proposed 1 | **63.6** | **59.8** | **58.5** |
| Noise 2 (babble) | Basic 1 | 61.6 | 59.9 | 56.7 |
| | Basic 2 | 60.3 | 55.4 | 52.1 |
| | Basic 3 | 59.2 | 57.8 | 55.4 |
| | Basic 4 | 61.7 | 57.8 | 56.7 |
| | Basic 5 | 55.6 | 54.0 | 53.5 |
| | Basic 6 | 58.0 | 57.6 | 57.4 |
| | Proposed 1 | **63.4** | **61.0** | **58.9** |
| Noise 3 (car) | Basic 1 | 62.5 | 59.9 | 57.0 |
| | Basic 2 | 60.9 | 56.3 | 54.3 |
| | Basic 3 | 60.6 | 56.8 | 56.6 |
| | Basic 4 | 60.5 | 57.5 | 55.5 |
| | Basic 5 | 56.9 | 55.1 | 53.4 |
| | Basic 6 | 58.2 | 57.8 | 57.5 |
| | Proposed 1 | **64.5** | **61.8** | **59.1** |
| Noise 4 (exhibition) | Basic 1 | 62.3 | 58.0 | 55.9 |
| | Basic 2 | 61.0 | 57.5 | 52.0 |
| | Basic 3 | 60.7 | 58.7 | 57.0 |
| | Basic 4 | 59.8 | 58.6 | 57.0 |
| | Basic 5 | 55.8 | 55.2 | 53.5 |
| | Basic 6 | 58.3 | 58.0 | 57.8 |
| | Proposed 1 | **66.0** | **61.5** | **60.6** |
| Noise 5 (restaurant) | Basic 1 | 61.4 | 58.6 | 56.3 |
| | Basic 2 | 58.3 | 55.0 | 52.1 |
| | Basic 3 | 60.3 | 56.7 | 55.4 |
| | Basic 4 | 61.4 | 58.9 | 56.3 |
| | Basic 5 | 55.8 | 55.0 | 53.2 |
| | Basic 6 | 57.9 | 57.8 | 57.6 |
| | Proposed 1 | **62.8** | **59.9** | **58.1** |
| Noise 6 (street) | Basic 1 | 62.5 | 58.6 | 56.9 |
| | Basic 2 | 60.2 | 56.4 | 53.5 |
| | Basic 3 | 61.9 | 58.3 | 57.0 |
| | Basic 4 | 60.2 | 57.4 | 56.0 |
| | Basic 5 | 55.7 | 54.0 | 53.7 |
| | Basic 6 | 58.0 | 57.9 | 57.5 |
| | Proposed 1 | **66.0** | **59.9** | **58.7** |
| Noise 7 (subway) | Basic 1 | 61.3 | 58.4 | 56.4 |
| | Basic 2 | 59.8 | 57.1 | 54.0 |
| | Basic 3 | 60.9 | 58.2 | 56.3 |
| | Basic 4 | 60.3 | 57.9 | 56.1 |
| | Basic 5 | 54.8 | 54.4 | 54.2 |
| | Basic 6 | 58.4 | 58.2 | 57.9 |
| | Proposed 1 | **65.7** | **61.0** | **58.6** |
| Noise 8 (train) | Basic 1 | 62.7 | 59.1 | 57.4 |
| | Basic 2 | 60.7 | 57.3 | 53.6 |
| | Basic 3 | 59.5 | 58.7 | 56.6 |
| | Basic 4 | 60.4 | 57.8 | 56.2 |
| | Basic 5 | 55.9 | 55.4 | 54.1 |
| | Basic 6 | 58.2 | 57.7 | 57.2 |
| | Proposed 1 | **64.2** | **60.7** | **58.0** |
| Average | Basic 1 | 62.0 | 59.0 | 56.7 |
| | Basic 2 | 60.1 | 56.3 | 53.3 |
| | Basic 3 | 60.5 | 58.0 | 56.4 |
| | Basic 4 | 60.8 | 58.1 | 56.2 |
| | Basic 5 | 55.9 | 54.8 | 53.8 |
| | Basic 6 | 58.1 | 57.8 | 57.5 |
| | Proposed 1 | **64.5** | **60.7** | **58.8** |

Table 2 shows that the proposed system outperforms all the basic systems in all the 8 noises along the different SNR levels. Also, it can be observed that although PLP+BFCC ("Basic 4") provides the second best results in some types of noise at high SNR, overall, the system with MFCC only ("Basic 1") produces the second best result in average.

B. Second proposed system

Although the first proposed system works better than all other ones, its performance still faces high degradation in the noisy case. Therefore, another study has been carried out by increasing the dimensions of the first two feature sets (MFCC and BFCC) from 10 to 23. In these conditions, PLP can be removed from the feature extraction module without a significant loss in performance. Fig 12. shows the diagram of the second proposed system.

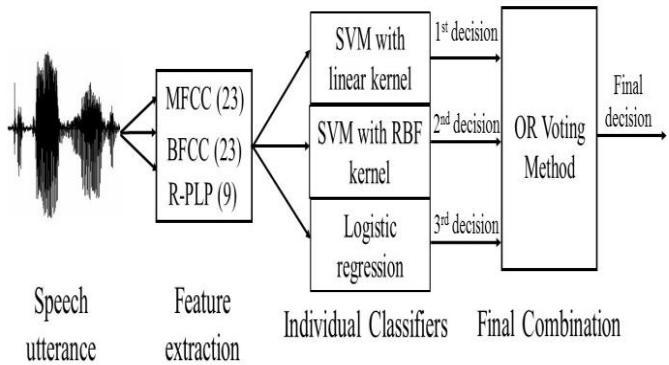

Fig. 12. The second proposed speaker verification system.

The average results over all the noise types and SNRs achieved by this system are shown in Fig. 13. Average AUC values for the three first MFCC-based basic systems are also shown for comparison purposes. For clarity, results with the PLP-based basic systems are not included in this figure. As it can be observed, in this case, the second proposed system performs much better than the basic ones. Also, this system presents an improvement around 4% or 5% absolute in comparison to the first proposed one. Although the results obtained by the second system are higher than the other systems, it still has a significant degradation caused by the presence of noise. In next subsection, a noise removal technique is used to modify the second proposed system in order to get better performance.

### C. Third proposed system

In this case, a noise removal technique is used as a preprocessing stage (before the feature extraction process) of the second proposed SV system with the aim of improving its performance in the presence of noise. After trying different techniques of noise removal such as Berouti's Spectral Subtraction, Minimum Mean-Square Error Short-Time Spectral Amplitude (MMSE-STSA), and Multiband Noise Removal, the latter one showed the highest performance.

This technique was proposed in 2002 by Kamath and Loizou [54]. Instead of subtracting the noise spectrum estimate over the entire speech spectrum (as in the case of the standard Spectral Subtraction), the multiband technique firstly divides the spectrum of the utterance into non-overlapping bands and then the subtraction is done in each band independently. This method works much better than the other ones because the real-world noises are colored ("not white").

The average results of the third proposed system over all the noise types and SNRs are shown in Fig. 13. Average AUC values for the three MFCC-based basic systems and the second proposed one are also shown for comparison purposes.

It can be observed a significant improvement around 7% or 8% absolute in comparison to the second proposed system due to the inclusion of the multiband noise removal technique. Also, these results provide an average improvement in AUC values around 15% absolute compared to the best basic system ("Basic 1").

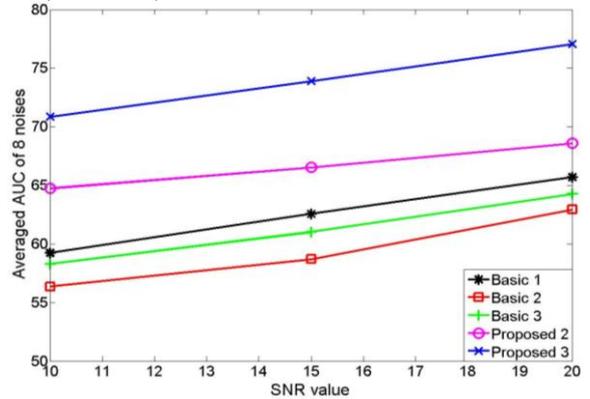

Fig. 13. Averaged AUC values overall noise types and SNRs with and without the multiband noise removal technique.

### VII. EXECUTION TIME

The previous two Sections have shown the success of the three proposed systems from the point of view of the performance. There is a tradeoff between the performance of the system and its execution time which is related to its computational complexity. This Section is dedicated to comparing the complexity of the three proposed systems and the three MFCC-based basic systems (with low and high feature dimensions) in terms of execution time. Although it can be predicted that the proposed systems will require more time because of the combination of features and classifiers, it is important to check if these systems can be applied in real-time speaker verification applications or not.

For this purpose, a whole experiment has been carried out using the same computer to ensure a fair comparison. The used one is Dell equipped with an Intel core 2 duo CPU at 2.0GHz and 2GB of RAM. Table 3 shows the averaged execution time for testing for all the systems. The execution time is the average over 22 speakers and 2 different utterances and it refers to the required time for doing the whole test phase: read the sound file, extract the feature vectors, and do the classification stage.

Table 3. Execution time for the basic and proposed systems averaged over 44 different testing utterances.

| Feature dimension | System | Execution time (in seconds) |
|---|---|---|
| Low | Basic 1 | 0.12 |
| | Basic 2 | 0.15 |
| | Basic 3 | 0.78 |
| | Proposed 1 | 0.76 |
| High | Basic 1 | 0.18 |
| | Basic 2 | 0.22 |
| | Basic 3 | 0.99 |
| | Proposed 2 | 0.88 |
| High with multiband noise removal | Proposed 3 | 1.86 |

From Table 3 we can draw the following remarks:

1. The systems which use MFCC only with linear SVM ("Basic 1") have the lowest execution time in comparison to any other systems regardless the dimension of the features.

2. The combination of all features ("Basic 3") dramatically increases the complexity of the system. This increment also has not positive effect on the AUC values as shown in Table 2 and Fig. 13.

3. Although the proposed systems 1 and 2 use a combination of three classifiers, they have approximately the same execution time as the Basic 3 system. This is due to the reduction of time in the feature extraction stage.

4. The multiband noise removal technique increases the execution time of the third proposed system, in such a way that it is the double of the execution time of the second proposed system. Nevertheless, it is worth noting that the noise removal function has not been optimized for speed.

5. Although the proposed systems require more processing time, this is still small (less than 2 sec.) so these systems can be used in real-time applications.

## VIII. Conclusions and Future Work

In this paper, a text-independent speaker recognition system for the verification task is proposed. The system is based on the combination of different acoustic features (MFCC, BFCC, PLP, and RASTA-PLP) to avoid the weak points of using each one individually. Besides, to achieve the optimal performance of the system, the combination of three classifiers (linear SVM, RBF SVM, and LR) is proposed in order to increase the generalization and learning abilities of the single classifiers. The combination of the classifiers is performed through a parallel structure using the OR voting rule method. The experiments were done firstly using clean speech. In this case, the performance of the system was very high (AUC around 98%). Secondly, the system was tested in noisy conditions with eight different types of noises and three SNR values. These experiments showed the success of the proposed system in comparison to other six basic ones. To enhance the results more, a second system using MFCC, BFCC and RASTA-PLP with higher dimensionality and the combination of classifiers is also presented. The performance of this second proposed system was better than either the basic systems or the first proposed one in noisy conditions.

Finally, in order to achieve better results in the presence of noise, the inclusion of a multiband noise removal technique as a preprocessing stage to the second proposed system was done. With this modification, an important improvement in the performance of the whole system was attained for all the noise types at all the SNR levels considered. The average improvement was more than 15% absolute in comparison to the basic systems with low feature dimensionality and around 7%-8% absolute compared to the second proposed one. This system could be used efficiently in real-time applications because it presents high performance together to an acceptable processing time.

Future work will be directed towards trying to optimize this system for reducing its execution time, especially in the noise removal stage. After that, the application of the proposed system to speaker identification tasks will be addressed.

## Acknowledgements

The authors want to thank Erasmus Mundus "Green-IT" program for its grant for providing the funding for this work. This work has also been partially supported by the Spanish Government Grant TEC2014-53390-P and by the Regional Government of Madrid S2013/ICE-2845-CASI-CAM–CM project. Thanks to the anonymous editor and the reviewers of Neural Computing and Applications journal for their constructive comments and suggestions, the final publication is available at https://link.springer.com/article/10.1007%2Fs00521-016-2470-x

## References

[1] M. F. Zanuy, and E. M. Moreno, "State of the art in speaker recognition", IEEE Aerospace & Electronic Systems Magazine, vol. 20, no. 5, pp. 7-12, May 2005.

[2] M. Hébert, "Text-dependent speaker recognition", In Springer handbook of speech processing, Springer Berlin Heidelberg, pp 743-762, 2008.

[3] H. Gish and M. Schmidt, "Text-independent speaker identification", IEEE Signal Processing Magazine, vol. 11, no. 4, pp. 18-32, October 1994.

[4] T. Kinnunen, and H. Li, "An overview of text-independent speaker recognition: From features to supervectors", Speech Communication, vol. 52, no. 1, pp.12-40, 2010.

[5] W. Yutai, L. Bo, J. Xiaoqing, L. Feng, and W. Lihao, "Speaker recognition based on dynamic MFCC parameters", International Conference on Image Analysis and Signal Processing (IASP), Taizhou, pp. 406-409, 11-12 April 2009.

[6] M. G. Sumithra, and A. K. Devika, "A study on feature extraction techniques for text independent speaker identification", International Conference on in Computer Communication and Informatics (ICCCI), Coimbatore, pp. 1-5, 10-12 January 2012.

[7] E. Ambikairajah, "Emerging features for speaker recognition", 6th International IEEE Conference on Information, Communications & Signal Processing, Singapore, pp. 1-7, 10-13 December 2007.

[8] J. P. Campbell, D. A. Reynolds and R. B. Dunn, "Fusing high-and low-level features for speaker recognition", in Proc. of European Conference on Speech Communication and Technology (EUROSPEECH), Geneva, Switzerland, pp. 2665-2668, September 2003.

[9] N. P. Jawarkar, R. S. Holambe, and T. K. Basu, "On the use of classifiers for text-independent speaker identification", First International Conference on Automation, Control, Energy and Systems (ACES), Hooghy, pp. 1-6, 1-2 February 2014.

[10] S. Parveen, A. Qadeer and P. Green, "Speaker recognition with recurrent neural networks", 6th International Conference on Spoken Language Processing (INTERSPEECH), Beijing, China, 16-20 October 2000.

[11] N. Almaadeed, A. Aggoun, and A. Amira, "Speaker identification using multimodal neural networks and wavelet analysis", in IET Biometrics, vol. 4, no. 1, pp.18-28, March 2015.

[12] D. Reynolds, T. Quatieri, and R. Dunn, "Speaker verification using adapted Gaussian mixture models", Digital Signal Processing, vol. 10, no. 1, pp. 19-41, 2000.

[13] S. Yaman, and J. Pelecanos, "Using Polynomial Kernel Support Vector Machines for Speaker Verification", IEEE Signal Processing Letters, vol. 20, no. 9, pp. 901-904, 11 July 2013.

[14] R. Solera-Ureña, J. Padrell-Sendra, D. Martin-Iglesias, A. Gallardo-Antolin, C. Pelaez-Moreno, and F. Diaz-de-Maria, "ch. SVMs for Automatic Speech Recognition: A Survey", Progress in Nonlinear Speech Processing, ser. Lecture Notes in Computer Science, Berlin, Heidelberg, Germany, Springer-Verlag, Vol. 4391, pp. 190–216, May 2007.

[15] R. Dehak, N. Dehak, P. Kenny and P. Dumouchel, "Kernel combination for SVM speaker verification", Speaker and Language Recognition Workshop (Odyssey 2008), Stellenbosch, South Africa, 21-24 January 2008.

[16] S. Farah, and A. Shamim, "Speaker recognition system using Mel-frequency cepstrum coefficients, linear prediction coding and vector quantization", 3rd International Conference on Computer, Control & Communication (IC4), Karachi, pp. 1-5, 25-26 September 2013.

[17] T. S. Gaafar, H. M. Abo Bakr, and M. I. Abdalla, "An improved method for speech/speaker recognition", International Conference on Informatics, Electronics & Vision (ICIEV), Dhaka, pp. 1-5, 23-24 May 2014.

[18] H. Maged, A. Abou El-Farag, and S. Mesbah, "Improving speaker identification system using discrete wavelet transform and AWGN", 5th IEEE International Conference on Software Engineering and Service Science (ICSESS), Beijing, pp. 1171-1176, 27-29 June 2014.

[19] W. M. Campbell, J. P. Campbell, D. A. Reynolds, E. Singer, and P. A. Torres-Carrasquillo, "Support vector machines for speaker and language recognition", Computer Speech and Language, vol. 20, no. 2, pp. 210-229, July 2006.

[20] J. C. Wang, L. X. Lian, Y. Y. Lin, and J. H. Zhao, "VLSI design for SVM-based speaker verification system", IEEE Transactions on Very Large Scale Integration (VLSI) Systems, vol. 23, no. 7, pp. 1355-1359, July 2015.

[21] A. Alarifi, I. Alkurtass, and A. S. Alsalman, "SVM based Arabic Speaker Verification System for Mobile Devices", International Conference on Information Technology and e-Services, Sousse, pp. 1-6, 24-26 March 2012.

[22] M. Wozniak, M. Grana, and E. Corchado, "A survey of multiple classifier systems as hybrid systems", Information Fusion, vol. 16, pp. 3-17, 2014.

[23] A. R. WEBB, "Statistical pattern recognition", John Wiley & Sons, 2003.

[24] J Kittler, M. Hatef, R. P. W. Duin, and J. Matas, "On Combining Classifiers", IEEE Transactions on Pattern Analysis and Machine Intelligence, vol. 20, no. 3, pp. 226-239, 1998.

[25] S. B. Davis, and P. Mermelstein, "Comparison of Parametric representations for Monosyllabic word recognition in continuously spoken sentences", IEEE Transactions on Acoustics, Speech, Signal Processing, vol. 28, pp. 357-366, August 1980.

[26] M. Cutajar, E. Gatt, I. Grech, O. Casha and J. Micallef, "Comparative study of automatic speech recognition techniques", IET Signal Processing, vol. 7, no. 1, pp. 25-46, February 2013.

[27] U. Sharma, S. Maheshkar, and A. N. Mishra, "Study of robust feature extraction techniques for speech recognition system", International Conference on Futuristic Trends on Computational Analysis and Knowledge Management (ABLAZE), Noida, pp. 654-658, February 2015.

[28] H. Hermansky, "Perceptual linear predictive (PLP) analysis of speech", Journal of the Acoustical Society of America (JASA), vol. 87, pp.1738–1752, April 1990.

[29] N. Dave1, "Feature extraction methods LPC, PLP and MFCC in speech recognition", International Journal for Advance Research in Engineering and Technology, vol. 1, no. 6, July 2013.

[30] M. Sahidullah, S. Chakroborty, and G. Saha, "On the use of perceptual Line Spectral Pairs Frequencies for speaker identification", International Journal of Biometrics, vol. 2, no. 4, pp. 358-378, 2010.

[31] H. Hermansky and N. Morgan, "RASTA Processing of Speech", IEEE Transactions On Speech and Audio Processing, vol. 2, pp. 578-589, Oct. 1994.

[32] A. J. Smola, and B. Schölkopf, "A tutorial on support vector regression", Journal of Statistics and computing, vol. 14, no. 3, pp.199-222, 2004.

[33] G. R. G. Lanckriet, N. Cristianini, P. Bartlett, L. E. Ghaoui, and M. I. Jordan, "Learning the Kernel Matrix with Semidefinite Programming", Journal of Machine Learning Reasearch, Vol. 5, pp. 27–72, 2004.

[34] C. J. Burges, "A tutorial on support vector machines for pattern recognition", Journal of Data mining and knowledge discovery", vol. 2, no. 2, pp.121-167, 1998.

[35] P. McCullagh, and J. A. Nelder, "Generalized linear models", vol. 37, CRC press, 1989.

[36] A. J. Dobson, "An introduction to generalized linear models", University of Newcastle, New South Wales, Australia. Chapman & Hall Ltd., London, 1990.

[37] D. W. Hosmer JR, and S. lemeshow, "Applied Logistic Regression", John Wiley & Sons, 2004.

[38] L. Feng, and L. K. Hansen, "A new database for speaker recognition", IMM, Informatics and Mathematical Modelling, DTU, 2005.

[39] http://www.pbs.org/wgbh/nova/pyramid [Online], last accessed: 2016-03-01.

[40] C. Micheloni, S. Canazza, and G. L. Foresti, "Audio–video biometric recognition for non-collaborative access granting", Journal of Visual Languages and Computing, vol. 20, no. 6, pp. 353-367, 2009.

[41] T. May, S. van de Par, and A. Kohlrausch, "Noise-Robust Speaker Recognition Combining Missing Data Techniques and Universal Background Modeling", IEEE Transactions on Audio, Speech and Language Processing, vol. 20, no. 1, pp. 108-121, 2012.

[42] A. P. Bradley, "The use of the area under the ROC curve in the evaluation of machine learning algorithms", Pattern Recognition, vol. 30, no. 7, pp. 1145–1159, July 1997.

[43] L. P. García-Perera, B. Raj, and J. A. Nolazco Flores, "Optimization of the DET curve in speaker verification under noisy conditions", 2013 IEEE International Conference on Acoustics, Speech and Signal Processing (ICASSP), pp. 7765-7769, Vancouver, Canada, 26-31 May 2013.

[44] M. Markaki, and Y. Stylianou, "Voice Pathology Detection and Discrimination Based on Modulation Spectral Features", IEEE Transactions on Audio, Speech, and Language Processing, vol. 19, no. 7, pp. 1938-1948, Sept. 2011.

[45] L. Uzan, and L. Wolf, "I know that voice: Identifying the voice actor behind the voice", International Conference on Biometrics (ICB), pp. 46-51, May 2015.

[46] Y. Lan, Z. Hu, Y. C. Soh, and G. B. Huang, "An extreme learning machine approach for speaker recognition", Journal of Neural Computing and Applications, vol. 22, no. 3-4, pp. 417-425, 2013.

[47] D. P. W. Ellis, "PLP and RASTA (and MFCC, and inversion) in Matlab", [Online]. Available: http://www.ee.columbia.edu/~dpwe/resources/matlab/rastamat/, last accessed: 2016-03-01.

[48] M. Zeppelzauer, "Discrimination and Retrieval of Animal Sounds", Master Dissertation, 2005.

[49] J. P. Openshaw, Z. P. Sun, and J. S. Mason, "A comparison of composite features under degraded speech in speaker recognition", IEEE International Conference on Acoustics, Speech, and Signal Processing (ICASSP-93), vol. 2, pp. 371-374, 1993.

[50] S. Furui, "Speaker-independent isolated word recognition using dynamic features of speech spectrum", IEEE Transactions on Acoustics, Speech and Signal Processing, vol. 34, no. 1, pp. 52-59, Feb. 1986.

[51] H. Aronowitz, "Unsupervised Compensation of Intra-Session Intra-Speaker Variability for Speaker Diarization", Speaker and Language Recognition Workshop (Odyssey 2010), pp. 138-145, Brno, Czech Republic, 28 June - 1 July 2010.

[52] L. M. Mazaira-Fernández, A. Álvarez-Marquina, and P. Gómez-Vilda, "Improving Speaker Recognition by Biometric Voice Deconstruction", Frontiers in Bioengineering and Biotechnology, vol. 3, 2015.

[53] S. V. Chougule, and M. S. Chavan, "Robust Spectral Features for Automatic Speaker Recognition in Mismatch Condition", Procedia Computer Science, vol. 58, pp. 272-279, 2015.

[54] S. D. Kamath and P. C. Loizou, "A multiband spectral subtraction method for enhancing speech corrupted by colored noise", IEEE International Conference on Acoustics, Speech, and Signal Processing (ICASSP), USA, Vol. 4, pp. 4164-4167, May 2002.